\documentstyle{article} 
\begin{document}

\title{Dynamic wormholes, anti-trapped surfaces, and energy conditions}
\author{
David Hochberg$^{+}$ and Matt Visser$^{++}$\\
\\
$^{+}$Laboratorio de Astrof\'{\i}sica Espacial y F\'{\i}sica Fundamental\\
Apartado 50727, 28080 Madrid, Spain\\
\\
$^{++}$Physics Department, Washington University\\
Saint Louis, Missouri 63130-4899, USA\\
}
\maketitle
\begin{abstract}
It is by now apparent that topology is too crude a tool to accurately
characterize a generic traversable wormhole.  In two earlier papers
we developed a complete characterization of generic but static
traversable wormholes, and in the present paper extend the discussion
to arbitrary time-dependent (dynamical) wormholes. A local definition
of wormhole throat, free from assumptions about asymptotic flatness,
symmetries, future and past null infinities, embedding diagrams,
topology, and even time-dependence is developed that accurately
captures the essence of what a wormhole throat is, and where it is
located. Adapting and extending a suggestion due to Page, we define
a wormhole throat to be a marginally anti-trapped surface, that
is, a closed two-dimensional spatial hypersurface such that one of
the two future-directed null geodesic congruences orthogonal to it
is just beginning to diverge.  Typically a dynamic wormhole will
possess {\em two} such throats, corresponding to the two orthogonal
null geodesic congruences, and these two throats will not coincide,
(though they do coalesce into a single throat in the static limit).
The divergence property of the null geodesics at the marginally
anti-trapped surface generalizes the ``flare-out'' condition for
an arbitrary wormhole.  We derive theorems regarding violations of
the null energy condition (NEC) at and near these throats and find
that, even for wormholes with arbitrary time-dependence, the
violation of the NEC is a generic property of wormhole throats. We
also discuss wormhole throats in the presence of fully antisymmetric
torsion and find that the energy condition violations {\em cannot}
be dumped into the torsion degrees of freedom. Finally by means of
a concrete example we demonstrate that even temporary suspension
of energy-condition violations is incompatible with the flare-out
property of dynamic throats.
\end{abstract}

\bigskip


\bigskip

e-mail: hochberg@laeff.esa.es; visser@kiwi.wustl.edu

\newpage

\renewcommand
\baselinestretch{1.5}

\section{Introduction}

Traversable Lorentzian wormholes~\cite{Morris-Thorne,MTY,Visser-Book}
have often been viewed as intrinsically topological objects,
with the topological nature of their spatial sections revealed
graphically by means of embedding diagrams and ``shape'' functions
as either ``handles'' in spacetime (intra-Universe wormholes
joining two distant regions of the same Universe) or as ``bridges''
(inter-Universe wormholes linking two distinct spacetimes ).
Both of these types of wormhole give rise to the notion of
multiply-connected Universes and spatio-temporal networks possessing
a non-trivial topology~\cite{Hochberg-Kephart}. More often than not,
global geometric constraints are imposed on the wormhole, as well
as symmetry properties.  For example, the static Morris-Thorne
inter-Universe wormhole is an example of this more restrictive
class in that it requires exact spherical symmetry and the existence
of two asymptotically flat regions in spacetime~\cite{Morris-Thorne}.
As we have previously argued~\cite{Hochberg-Visser,Visser-Hochberg}
there are many other spacetime configurations and geometries that
one might still quite reasonably want to classify as wormholes,
that either do not possess any asymptotically flat regions, or have
trivial topology, or exhibit both these features. An example of
the former is provided by the Hochberg-Popov-Sushkov self-consistent
semi-classical wormhole (which is a wormhole of the inter-Universe
type joining up two spacetimes with no asymptotically flat spatial
regions)~\cite{HPS}.  Examples of the topologically trivial
wormholes~\cite{Visser-Book} are provided by a closed
Friedmann-Robertson-Walker (FRW) spacetime joined to an ordinary
Minkowski spacetime by a narrow neck or two closed FRW spacetimes
joined by a bridge~\cite{Hochberg-Visser,Visser-Hochberg}.

The only difference between these two classes of wormholes ({\em
i.e.} bridges and handles {\em versus} topologically trivial) arises
at the level of {\em global} geometry and {\em global} topology.
This suggests that it is important to identify a fundamental {\em
local} property that can be used to characterize what one means by
a wormhole, an intrinsic property to be abstracted from the broad
phylum of wormholes which can then be used to unambiguously define
what is meant by a wormhole.  Indeed, the local physics, that which
is operative near the ``throat'' of the wormhole, is insensitive
to global properties and indicates that a local definition of what
is meant by a wormhole throat is called for.  This definition should
be based solely on local properties and be free from technical
assumptions about asymptotic flatness, future and past null
infinities, global hyperbolicity, symmetries, embeddings and
topology.

In two previous papers~\cite{Hochberg-Visser,Visser-Hochberg} we
have performed such an analysis for static traversable wormholes.
In this paper, we lift the static restriction and shall investigate
the generic (not necessarily static) traversable wormhole.  We make
no assumptions about symmetries, spherical or otherwise, nor do we
assume the existence of asymptotically flat regions. To proceed,
we first have to define exactly what we mean by a wormhole and we
find, just as in the treatment of the generic {\em static}
case~\cite{Hochberg-Visser,Visser-Hochberg}, that there is a natural
local {\em geometric} (not topological) characterization of the
existence and location of a wormhole ``throat''. This characterization
is developed in the language of the expansion of null geodesic
congruences propagating outward from, and orthogonal to, closed
two-dimensional spatial hypersurfaces (denoted $\Sigma$).  The
congruence is subject to a ``flare-out'' condition that suitably
generalizes that of the Morris-Thorne analysis. But, unlike that
earlier definition~\cite{Morris-Thorne,MTY}, ours makes no reference
to embeddings or shape-functions.  In this language, the spatial
hypersurface in question will be a wormhole throat provided the
expansion $\theta_{\pm}$ of one of the two orthogonal null congruences
vanishes on that surface: $\theta_+ = 0$ and/or $\theta_{-} = 0$,
and if the rate-of-change of the expansion along the {\em same}
null direction ($u_\pm$) is positive-semi-definite at the surface:
$d \theta_\pm/d u_\pm \geq 0$. This latter constraint is precisely
the ``flare-out'' condition generalized to an arbitrary wormhole.
These two conditions on the expansion define the throat to be a
minimal hypersurface, {\em i.e.}, an extremal surface of minimal
area (with respect to deformations in the appropriate $u_\pm$ null
direction).  Thus, a wormhole throat is a {\em marginally anti-trapped
surface}.  Historically, Page~\cite{Page} was the first to suggest
that under suitable circumstances a wormhole throat could be viewed
as an anti-trapped surface in spacetime, and we shall soon see that
this definition promises to be the most efficient and most
physical framework for generalizing the concept of throat to the
fully arbitrary and dynamic case.

While this definition captures the intuitive concept of throat
admirably, there can be cases calling for slight definitional
refinements, for example, when $d \theta_{\pm}/d u_{\pm} > 0$ is
strictly positive on the throat, in which case we are dealing with
a {\em strongly anti-trapped surface}, as well as other cases for
which weaker, averaged notions of flare-out will suffice.

In general, the vanishing of the independent expansions $\theta_+
= 0$ and $\theta_- = 0$ will take place on two distinct hypersurfaces.
Thus, (dynamical) wormholes generally possess {\em two} throats
provided each hypersurface is individually flared-out: $d
\theta_+/du_+ \geq 0$ on $\Sigma_{u+}$, and $d \theta_-/du_- \geq
0$ on $\Sigma_{u-}$. Of course, the two throats must (and they
do) coincide in the static limit.

With these definitions in place, we move on to develop a number of
theorems about the existence of matter at or near the throat(s)
violating the null energy condition (NEC).  These theorems make
repeated use of the Raychaudhuri equation for the expansions
$\theta_{\pm}$.  These results are local and pointwise, in distinction
to energy conditions obtained by averaging over inextendible null
geodesics, which are global in nature.  These energy theorems
generalize the original Morris--Thorne result by demonstrating
unequivocally that the NEC is generically violated at some points
on or {\em near} the two-dimensional hypersurface comprising the
throat(s).  This is an important result since these theorems hold
for an arbitrary dynamic or static wormhole irrespective of symmetries
or other global concerns and demonstrate that the energy condition
violations are truly generic. Our results are (of course)
also completely in accord with  the topological censorship theorem
of Friedman, Schleich, and Witt~\cite{FSW}.

The striking nature of the violations of the null energy condition
first discovered for the Morris-Thorne
wormhole~\cite{Morris-Thorne,MTY,Visser-Book}, has led numerous
authors to try and find ways of evading or minimizing these
violations. Most of these attempts focus on alternative gravity
theories, be they Brans-Dicke, dilaton gravity, higher-derivative
theories, {\em etc}. What all these extensions of Einstein gravity
``accomplish'' from a practical point of view is to provide one
with additional degrees of freedom (beyond the metric), which under
certain circumstances can be coerced into absorbing the energy
condition violations (leaving the remaining ordinary-matter sector
free to satisfy the classical energy conditions). Nevertheless,
the total effective stress energy tensor will violate the null
energy condition at or near the throat, so sweeping the unavoidable
energy condition violations into a specific sector does not make
the problem go away. This important but oft overlooked point has
been treated in some detail in~\cite{Visser-Hochberg}. (We
would be remiss in not warning the reader that a sizable fraction
of the published papers claiming to build wormholes without violating
the energy conditions suffer from severe technical problems, and are
often internally inconsistent.)

Similar comments apply of course to gravity plus torsion, although
theories with torsion are distinguished from other variants of
gravity by the fact that non-zero torsion gives rise to a non-trivial
contribution to the Raychaudhuri equation which {\em cannot} be
absorbed into an effective total stress energy tensor.  Moreover,
torsion appears naturally (and unavoidably) in theories of gravity
based on low-energy closed string theories.  These facts
make it of interest to treat the torsion case separately and in
some detail to assess the ability of torsion to defocus (null)
geodesics and to check the status of the NEC for throats in the
presence of torsion.  We find that totally antisymmetric torsion
actually {\em promotes} the energy condition violation at the throat
(but helps to lessen it away from the throat by generating twist).
Other attempts to get around the energy-condition violations have
led to considerations of time-dependent wormholes. In this domain,
it is indeed possible to temporarily suspend the violations, but
only at the heavy expense of totally destroying the flare-out
properties of the throat.

Since the Raychaudhuri equation with torsion is not standard textbook
fare, we include a brief resum\'e of torsion in Section II to
establish the notation used in the rest of the paper and provide
a simple derivation of the generalized Raychaudhuri and the companion
twist equations corresponding to the two independent null
congruences in Section III. We then define wormhole throats in
terms of the expansions in Section IV and prove the coalescence of
the two throats in the static limit.  Armed with these definitions,
we go on to derive the energy condition theorems for wormholes in
normal spacetime as well as in the presence of torsion in Section
V. Worked examples of dynamic wormholes are provided in Section
VI where, among other things, we show how the temporal suspension
of energy-condition violations eradicates the throat. Conclusions
and a discussion of our results are collected in Section VII.

\section{\protect{Geometric preliminaries:
spacetimes with \hfil\break torsion}}

In preparation for the derivation of the Raychaudhuri equation
governing the expansion in the presence of torsion, and to establish
the notation to be used throughout, we collect here a few basic
definitions and identities which will prove useful later on.
(Basic definitions regarding torsion can be gleaned
from~\cite{deFelice-Clarke,EGH,Schouten}, while an overview of
torsion in the string theory context can be extracted from~\cite{GSW}.)

As we are interested in keeping our discussion as general as
possible, we endeavour to work in a coordinate-free language and
to this end, shall make use of the abstract index notation, later
specializing when and if needed, to explicit coordinate systems.
For the time being then, the use of lower case latin letters
designates abstract indices: $(a,b,c,...)$ and run from 0 to 3.
(See Wald~\cite{Wald} for a discussion of the subtleties
associated with the use of ``abstract indices''.)  Let $v_a$ be a
covariant vector, its covariant derivative is
\begin{equation}
\label{E:covariant}
\nabla_a v_b = \partial_a v_b - C^c_{ab} v_c,
\end{equation}
where $C^b_{ac}$ denotes the connection of the underlying
four-dimensional spacetime. In principle, the connection can be
any ``tensor'' field guaranteeing that the covariant derivative
(\ref{E:covariant}) based upon it satisfies all the usual properties
(linear, Leibnitz, etc.)~\cite{Wald}.  However, we will not impose
the torsion-free condition, which means that the (total)
connection can be decomposed as
\begin{equation}
\label{E:connection}
C^c_{ab} = \Gamma^{c}_{ab} + H^{c}_{ab},
\end{equation}
where $C^c_{(ab)}= \frac{1}{2}(C^c_{ab}+C^c_{ba}) = \Gamma^{c}_{ab}$
is the ordinary symmetric Christoffel connection, depending on the
metric in the usual way, while $C^c_{[ab]}= \frac{1}{2}(C^c_{ab}-
C^c_{ba}) = H^{c}_{ab}$ defines the torsion, which is manifestly
anti-symmetric in its two lower indices.

Due to the mixed-symmetry of the connection, the commutator of the 
covariant derivative, which is used to define the curvature tensor, 
works out to be
\begin{eqnarray}
\label{E:commutator}
[\nabla_a,\nabla_b]v_c &=& (-2\partial_{[a} C^d_{b]c} 
+ 2C^e_{[a|c}C^d_{b]e})v_d -2C^e_{[ab]} \nabla_e v_c\nonumber \\
&=& {\bar R_{ab,c}}{}^d(C) v_d -2H^e_{ab}\nabla_e v_c,
\end{eqnarray}
where
\begin{equation}
\label{E:curvature}
{\bar R_{ab,c}}{ }^d(C) = -2\partial_{[a} C^d_{b]c} 
+ 2C^e_{[a|c}C^d_{b]e},
\end{equation}
is the associated curvature tensor.  The vertical bar within the
antisymmetrization brackets indicates that one is to antisymmetrize
over the pair $a$ and $b$, but not $c$.  We have distinguished the
curvature with an overbar in order to emphasize that this tensor
is not the ordinary Riemann tensor, unless the torsion vanishes
identically.  It {\em is} however the curvature associated with a
general connection $C$.  We note that the derivative of a vector
couples directly to the torsion, as evidenced by the second term
in the above identity (\ref{E:commutator}). The torsion also shows up
explicitly (and implicitly in the covariant derivatives) in the
commutator of two vector fields:
\begin{equation}
\label{E:commute}
[v,w]^b = v^a\nabla_a w^b - w^a\nabla_a v^b - 2v^a w^c H^b_{ac}.
\end{equation}

Although (\ref{E:curvature}) is not the standard Riemann tensor, it
is related to it as follows:
\begin{equation}
\label{E:curvature2}
{\bar R_{ab,c}}{ }^d(C) = R_{ab,c}{ }^d(\Gamma) 
-(\tilde \nabla_a H^d_{bc}- \tilde \nabla_b H^d_{ac})
+ 2 H^e_{[a|c}H^d_{b]e},
\end{equation}
where the covariant derivatives of the torsion are calculated with
respect to the symmetric (Christoffel) part of the connection only;
that is,
\begin{equation}
\tilde \nabla_a H^d_{bc} = \partial_a  H^d_{bc} 
-\Gamma^e_{ab}H^d_{ec} -\Gamma^e_{ac}H^d_{be} + \Gamma^d_{ae}H^e_{bc}.
\end{equation}
This identity (\ref{E:curvature2}) suggests that the torsion can be
regarded as a dynamic field propagating over a normal Riemannian
spacetime, {\em i.e.}, may either be regarded as fundamentally geometric,
as part and parcel of the connection (\ref{E:connection}), or as a
``matter'' tensor field in a spacetime with a conventional symmetric
connection.  We can make this latter association more precise by
writing the action from which we will infer the corresponding
equations of motion. We form the equivalent of the Einstein-Hilbert
action for the generalized curvature and allow for the presence of
ordinary matter (every other dynamical field imaginable except for
the metric and torsion):
\begin{equation}
\label{E:action}
S = -\frac{1}{16\pi} \int d^4x\, \sqrt{-g}\, \bar R(C)
+  \int d^4x\, \sqrt{-g} \, {\cal L}_{matter},
\end{equation}
where the generalized scalar curvature is $\bar R(C) = g^{ac} \bar
R_{ab,c}{ }^b (C)$ and is related to the scalar of Riemannian
curvature via
\begin{equation}
\label{E:scalar}
\bar R(C) = R(\Gamma) - g^{bc}\tilde \nabla_b H^a_{{}a c}
- H_{abc}H^{abc},
\end{equation}
which follows from (\ref{E:curvature2}) and using the covariant
constancy of the metric $\tilde \nabla_a g_{bc} = 0$, with respect
to $\tilde \nabla$. (Mathematically, it is possible to consider
even more general affine connections for which the covariant
derivative of the metric is not zero. The most general such affine
connection is then a linear combination of the Christoffel connection,
the torsion tensor, and a ``non-metricality tensor''. We will not
generalise our analysis to this level of abstraction as little
seems to be gained, and there are good physics reasons for keeping
the covariant derivative of the metric zero.)

Thus far, we have kept the treatment of the torsion part of the
connection completely general. If we now identify the torsion with
the totally anti-symmetric rank-three field strength ${\bf H} =
d{\bf A}$, where $\bf A$ is a two-form potential, or in terms of
components
\begin{equation}
\label{E:kalb}
H_{abc} = \partial_a A_{bc} + \partial_b A_{ca} + \partial_c A_{ab},
\end{equation}
then we have an explicit realization of torsion that is known to
arise naturally in closed string-theoretic low energy
gravity~\cite{GSW,Scherk,Nepomechie}.  In this particular incarnation
as an antisymmetric rank-three tensor, the torsion is also known
as the Kalb-Ramond field.  From here on, when we refer to torsion,
it will be of this form.

The equations of motion now follow immediately upon varying the
full action (\ref{E:action}) with respect to the metric, torsion, and
what ever other matter fields may be present.  The equation of
motion for the metric is given by
\begin{equation}
\label{E:Einstein}
G_{ab}(\Gamma) = \left(R_{ab}-\frac{1}{2}g_{ab}R \right) =
8\pi T_{ab} + 3H_{ade} H_b^{de} -\frac{1}{2}g_{ab} \; H_{cde}H^{cde},
\end{equation}
where $T_{ab}$ is the complete stress-energy tensor for the matter
fields. We see that although $H$ originates from the connection,
it can also be treated as simply an additional species of matter
and can therefore be shifted into an effective matter stress tensor.
However, it is of more than academic interest not to do so at this
stage. When we come to consider the expansion and twist of (null)
geodesic congruences in spacetimes with torsion, we will find that
the torsion makes explicit non-dynamic contributions to the
differential equations for the expansion and twist that cannot be
re-defined away, as it were, by invoking the equations of motion,
or by redefining the total effective stress energy tensor. Thus,
it will be of interest to see what influence the torsion may have
to focus and defocus bundles of null geodesics.  The equation of
motion for the torsion that follows from varying (\ref{E:action}) is
simply that
\begin{equation}
\tilde \nabla_a H^{abc}= 0.
\end{equation}
Using the metric equation (\ref{E:Einstein}), it follows that 
the Ricci tensor obeys the equation
\begin{equation}
\label{E:Ricci}
R_{ab} = 8\pi[T_{ab}-\frac{1}{2}g_{ab}T] + 3H_{ade}H_b^{de}
-g_{ab} \; (H_{ade}H^{ade}),
\end{equation}
while the scalar curvature is 
\begin{equation}
R =- 8\pi T - H_{ade}H^{ade}.
\end{equation}

\section{Null geodesic congruences}

We start by considering a compact two-dimensional hypersurface that
is both orientable and embedded into spacetime in a two-sided manner
in such a way that the induced two-metric is spacelike.  To discuss
the null geodesic congruences orthogonal to this surface, we shall,
following the description of Carter~\cite{Carter} begin by introducing
a future-directed ``outgoing'' null vector $l_{+}^a$, a future-directed
``ingoing'' null vector $l_{-}^a$ and a spatial orthogonal projection
tensor $\gamma^{ab}$ satisfying the following relations:
\begin{eqnarray}
\label{E:properties}
l^a_{+}l_{+a} &=& l^a_{-}l_{-a} = 0 \nonumber \\
l^a_{+}l_{-a} &=& l^a_{-}l_{+a} = -1 \nonumber \\
l^a_{\pm}\gamma_{ab} &=&  0\nonumber \\
\gamma^a_c\gamma^{cd} &=& \gamma^{ad}.
\end{eqnarray}
In terms of these null vectors and projector, we can decompose the
full spacetime metric (indeed, any tensor) uniquely:
\begin{equation}
\label{E:decomposition}
g_{ab}= \gamma_{ab} - l_{-a}l_{+b} - l_{+a}l_{-b}.
\end{equation}
Physically, this decomposition leads to a parameterization of
spacetime points in terms of two spatial coordinates (typically
denoted $x$) plus two null coordinates [$u_\pm$, or sometimes
$(u,v)$]. (We do not want to prejudice matters by taking the
words ``outgoing'' and ``ingoing'' too literally, since outside and
inside do not necessarily make much sense in situations of nontrivial
topology. The critical issue is that the spacelike hypersurface
must have two sides and $+$ and $-$ are just two convenient labels
for the two null directions.)

We consider the tensor fields defined by the covariant derivative of
the future-directed null vectors (there is one such tensor field for each
null congruence)
\begin{equation}
\label{E:B}
B^{\pm}_{ab} \equiv \nabla_b l_{\pm a},
\end{equation}
and ask for their rate of change along the corresponding null geodesic
parameterized with affine parameter $u_{\pm}$:
\begin{eqnarray}
\label{E:master}
\frac{d B^{\pm}_{ab}}{d u_{\pm}} 
\equiv 
l^c_{\pm}  \; \nabla_c B^{\pm}_{ab} &=& 
l^c_{\pm}  \; \nabla_c\nabla_b l_{\pm a}
\nonumber \\
&=& 
l^c_{\pm}  \; \nabla_b\nabla_c l_{\pm a} + 
l^c_{\pm} [\nabla_c,\nabla_b]l_{\pm a}\nonumber \\
&=&
-\nabla_b l^c_{\pm} \; \nabla_c l_{\pm a} + 
l^c_{\pm}[\nabla_c,\nabla_b]l_{\pm a}\nonumber \\
&=& -{B^{\pm}}^c_b \; B^{\pm}_{ac} + 
\bar R_{cb,a}{}^d(C)l_{\pm d}l^c_{\pm} -2l^c_{\pm}H_{cb} B^{\pm}_{ae}.
\end{eqnarray}
This uses the fact that the parallel transport of a tangent vector
along its corresponding geodesic vanishes:  $l^c_{\pm}\nabla_c
l_{\pm b} = 0$ (see technical comment below dealing
with non-affine parameterizations), plus the commutator identity
in (\ref{E:commutator}). 

In contrast to the case of timelike geodesics, the tensor field
$B^{\pm}_{ab}$ is not purely spacelike but has in addition, mixed
null-spacelike components:
\begin{eqnarray}
\nabla_al_{+b} &=& \gamma_a^c \gamma_b^d \; \nabla_cl_{+d} 
-l_{+b} \; \gamma_a^d \; l^c_{-} \; \nabla_dl_{+c} \nonumber \\
&=& v^{+}_{ab} -l_{+b} \gamma_a^d \; l^c_{-} \; \nabla_dl_{+c},
\end{eqnarray}
and 
\begin{eqnarray}
\nabla_a l_{-b} &=& \gamma_a^c \gamma_b^d \; \nabla_c l_{-d} 
-l_{-b} \; \gamma_a^d \; l^c_{+} \; \nabla_d l_{-c} \nonumber \\
&=& v^{-}_{ab} -l_{-b} \gamma_a^d l^c_{+} \; \nabla_d l_{-c}.
\end{eqnarray}
which define the purely spatial tensors $v^{\pm}_{ab} = \gamma_a^c
\gamma_b^d \nabla_cl_{\pm d} $, which admit the further decomposition as
follows ($\gamma^{ab} \gamma_{ba} = 2$):
\begin{eqnarray}
\label{E:v}
v^{\pm}_{ab} &=& \frac{1}{2}\theta_{\pm}\gamma_{ab} + 
\sigma^{\pm}_{ab} + \omega^{\pm}_{ab} \\
\theta_{\pm} &=& \gamma^{ab}v^{\pm}_{ab} = g^{ab}\nabla_al_{\pm b} \\
\sigma^{\pm}_{ab} &=& v^{\pm}_{(ab)} - \frac{1}{2}\theta_{\pm}\gamma_{ab} \\
\omega^{\pm}_{ab} &=& v^{\pm}_{[ab]},
\end{eqnarray}
where $\theta_{\pm}$ is the trace of $v^{\pm}_{ab}$ and provides
the measure of the instantaneous expansion of the cross-sectional
area of a bundle of null geodesics, while $\sigma^{\pm}_{ab}$ and
$\omega^{\pm}_{ab}$ denote the shear and twist, respectively, and
are also purely spatial tensors.

{From} these relations one may derive rate-of-change equations for
the expansion, shear and twist with respect to the 
corresponding affine parameters
$u_{\pm}$ starting from (\ref{E:master}), though we shall be primarily
interested in the rate of change of the expansion $\theta_{\pm}$
as this equation
will play a fundamental role later on when we come to define a
generic wormhole throat. So, taking the trace of (\ref{E:master})
yields a generalized version  of the Raychaudhuri equation (generalized
as it contains the effects of torsion) for the two expansions
(one for the $(+)$-congruence, the other for the $(-)$-congruence):
\begin{eqnarray}
\label{E:Raychaudhuri}
\frac{d \theta_{\pm}}{d u_{\pm}}= -\frac{1}{2}{\theta_{\pm}}^2 &-& 
{\sigma^{\pm}}^{ab}\sigma_{\pm ab}
+ {\omega^{\pm}}^{ab}\omega_{\pm ab} 
- R_c^d(\Gamma)l^c_{\pm} l_{\pm d} \nonumber \\
&-& 2H^d_{cb} {B^{\pm}}^b_d l^c_{\pm} + H_{eac}H^{ead}l^c_{\pm} l_{\pm d}.
\end{eqnarray}
With a  view to applications for deriving the energy conditions
associated with generic wormhole throats, it is useful to have at
hand the companion equation governing the rate of change of the
twist along null geodesics. This is derived by going back to
(\ref{E:master}), antisymmetrizing on the free indices and projecting
out the purely spatial part of the resulting equation. These two
operations yield a generalization of the twist equation [again,
one for the $(+)$--congruence, the other for the $(-)$--congruence]:
\begin{eqnarray}
\label{E:vtwist}
\frac{d \omega^{\pm}_{ba}}{d u_{\pm}} = 
 -\theta_{\pm} \omega^{\pm}_{ba} - 2{\sigma^{\pm}}^c_{[a}\omega^{\pm}_{b]c}
+ \tilde \nabla_c H^d_{ab}l^c_{\pm}l_{\pm d}
&+& H^e_{c[a}H^d_{b]e}l^c_{\pm}l_{\pm d}\nonumber \\ 
- 2{l^{\pm}}^cH^e_{c[b}
{B^{\pm}}_{a]e}.
\end{eqnarray}
The term linear in $H$ that appears in both the expansion and twist
equations is purely geometrical in origin, arising as it does, from
the commutator of two torsion-bearing covariant derivatives
(\ref{E:commutator}).  The other torsion contributions are dynamic in
origin, as these arise instead directly from the action and equations
of motion.  These features distinguish the torsion from all other
fields. Of course, in the absence of torsion,
these reduce to the standard Raychaudhuri and twist equations, for
$\theta_{\pm}$ and $\omega^{\pm}$, respectively~\cite{Hawking-Ellis,Wald}.

Technical aside: if one is  working with a non-affine parameterization
for the null congruences, then the parallel transport equation
becomes $l^c_{\pm}\nabla_c l_{\pm b} = {\cal K}_{\pm}l_{\pm b}$
where ${\cal K}_{\pm} = -l^a_{\mp}l^b_{\pm}\nabla_bl_{\pm a}$.
The expansion is still given by the trace of the spatial part of
$\nabla_a l_{\pm b}$ and we have that $\theta_{\pm} =
\gamma^{ab}v^{\pm}_{ab} = g^{ab}\nabla_al_{\pm b} - {\cal K}_{\pm}$.
The Raychaudhuri equation (\ref{E:Raychaudhuri}) will then pick up an
extra factor of ${\cal K}_{\pm}\theta_{\pm}$~\cite{Carter}.

\section{Definition of generic wormhole throats}

Our aim is to provide a precise, local, and robust geometric
definition of a (traversable) wormhole throat, equally valid for
static as well as time-dependent wormholes.  As a guide, we recall
that in the generic but {\em static} case, the throat was defined
as a two-dimensional hypersurface of minimal
area~\cite{Hochberg-Visser,Visser-Hochberg}.  The time independence
allows one to locate that minimal hypersurface entirely within one
of the constant-time three-dimensional spatial slices, and the
conditions of extremality and minimality can be applied and enforced
within that single time-slice. For a static throat, variational
principles involve performing arbitrary time-independent surface
deformations of the hypersurface in the remaining spatial direction
orthogonal to the hypersurface, which can always be taken to be
locally Gaussian. By contrast, in the time-dependent case, it may
not be possible to locate the entire throat within one time slice,
as the dynamic throat is an extended object in spacetime, and the
variational principle must be carried out employing surface
deformations in the two independent $\em null$ directions orthogonal
to the hypersurface: say, $\delta u_+$ and $\delta u_{-}$.  This,
by the way, suggests why it is that the embedding of the spatial
part of a wormhole spacetime in an Euclidean ${\bf R}^3$ is no
longer a reliable operational technique for defining ``flare-out''
in the time-dependent case.  Of course, in the static limit these
two variations will no longer be independent and arbitrary deformations
in the two null directions reduce to a single variation in the
constant-time spatial direction (see below). Realizing that the
time-dependent wormhole typically has two non-coincident throats
was perhaps the major conceptual stumbling block to overcome in
developing this formalism.

\subsection{Preliminaries}

In the following, we set up and define the properties of throats
in terms of the null congruences. Bear in mind that a throat will
be characterized in terms of the behavior of a single set of null
geodesics orthogonal to it.  We define a wormhole throat $\Sigma_{u
+}$ (there is also one for the other null congruence) to be a closed
2-dimensional hypersurface of minimal area taken in one of the
constant-$u_{+}$ slices, where $u_{+}$ is an affine parameter
suitable for parameterizing the future-directed null geodesics
$l_+$ orthogonal to $\Sigma_{u +}$.   All this means is that we
imagine ``starting'' off a collection of light pulses along the
hypersurface and we can always arrange the affine parameterizations
of each pulse to be equal to some constant on the hypersurface; we
take this constant to be zero.  We wish to emphasize that there is
a corresponding definition for the other throat $\Sigma_{u-}$. In
the following, we define and develop the conditions that both
hypersurfaces must satisfy individually to be considered as throats,
and shall do so in a unified way by treating them together by
employing the $\pm$-label.  Our next task is to compute the
hypersurface areas and impose the conditions of extremality and
minimality directly and to express these constraints in terms of
the expansion of the null geodesics. The area of $\Sigma_{u\pm}$
is given by
\begin{equation}
\label{E:area-v}
A(\Sigma_{u\pm}) = \int_{\Sigma_{u\pm}} \sqrt{\gamma} \, d^2x.
\end{equation}

An arbitrary variation of the surface with respect to deformations
in the null direction parameterized by $u_{\pm}$ is 
\begin{eqnarray}
\label{E:firstvary}
\delta A (\Sigma_{u\pm}) & = & \int_{\Sigma_{u\pm}}
 \frac{d \sqrt{\gamma}}{d u_{\pm}}
\,\delta u_{\pm}(x) 
\, d^2x.\nonumber \\
&=& \int_{\Sigma_{u\pm}} \sqrt{\gamma} \, \frac{1}{2}\gamma^{ab}\,
\frac{d \gamma_{ab}}{d u_{\pm}} \, \delta u_{\pm}(x) 
\, d^2x.
\end{eqnarray}
If this is to vanish for arbitrary variations $\delta u_{\pm}(x)$, then
we must have that 
\begin{equation}
\frac{1}{2}\gamma^{ab}
\frac{d \gamma_{ab}}{d u_{\pm}} = 0,
\end{equation}
which expresses the fact that the hypersurface 
$\Sigma_{u \pm}$ is extremal.

This condition of hypersurface extremality can also be phrased
equivalently and directly in terms of the expansion of the null
congruences.  The simplest way to do so is to consider the Lie
derivative ${\cal L}^{\pm}_l$ acting on the full spacetime metric:
\begin{eqnarray}
{\cal L}^{\pm}_l g_{ab} 
&=& l^c_{\pm}\nabla_c g_{ab} + g_{cb}\nabla_al^c_{\pm} 
+ g_{ac}\nabla_b l^c_{\pm}\nonumber \\
&=& \nabla_a l_{\pm b} + \nabla_b l_{\pm a}\nonumber \\
&=& {B^{\pm}}_{ba} + {B^{\pm}}_{ab} = 2{B^{\pm}}_{(ab)},
\end{eqnarray}
with the second equality holding provided the metric is covariantly
constant with respect to the full covariant derivative, which is
in fact the case, even in the presence of arbitrary torsion.
We now use the decomposition (\ref{E:decomposition}) of the spacetime
metric and work out the Lie derivative using the Leibnitz rule:
\begin{eqnarray}
{B^{\pm}}_{(ab)} 
&=& \frac{1}{2}{\cal L}^{\pm}_l g_{ab} \nonumber \\
&=& \frac{1}{2}{\cal L}^{\pm}_l 
(\gamma_{ab} -l_{-a}l_{+ b} -l_{+ a} l_{- b}),\nonumber \\
&=& \frac{1}{2}{\cal L}^{\pm}_l \gamma_{ab} 
-\frac{1}{2}(l_{-a} {\cal L}^{\pm}_ll_{+b}
 + l_{+b} {\cal L}^{\pm}_l l_{-a} + (a \leftrightarrow b) ),
\end{eqnarray}
from which, and using the properties in (\ref{E:properties}), implies
\begin{eqnarray}
\theta_{\pm} &=&  g^{ab}{B^{\pm}}_{(ab)} =
\gamma^{ab}{B^{\pm}}_{(ab)} = \gamma^{ab}v^{\pm}_{ab}\nonumber \\
       &=& \frac{1}{2}\gamma^{ab}{\cal L}^{\pm}_l \gamma_{ab}\nonumber \\
       &=& \frac{1}{2}\gamma^{ab}\frac{d \gamma_{ab}}{d  u_{\pm}}.
\end{eqnarray}
So the condition that the area of the hypersurface be extremal is
simply that the expansion of the null geodesics vanish at the
surface: $\theta_{\pm} = 0$.  To ensure that the area be $\em minimal$,
we need to impose an additional constraint and shall require that
$\delta^2 A (\Sigma_{u\pm}) \geq 0$. By explicit computation,
\begin{eqnarray}
\label{E:minimality}
\delta^2 A(\Sigma_{u\pm}) &=& \int _{\Sigma_{u\pm}} \sqrt{\gamma} \left(
{\theta_{\pm}}^2 + \frac{d \theta_{\pm}}{d u_{\pm}} \right)
\delta u_{\pm}(x)\, \delta u_{\pm}(x) d^2x \nonumber \\
&=& \int_{\Sigma_{u\pm}} \sqrt{\gamma}\, \frac{d \theta_{\pm}}{d u_{\pm}}
\, \delta u_{\pm}(x)\, \delta u_{\pm}(x)\, d^2x \geq 0,
\end{eqnarray}
where we have used the extremality condition ($\theta_{\pm} = 0$) in
arriving at this last inequality.  For this to hold at the throat
for arbitrary variations $\delta u_{\pm}(x)$ , and 
since $(\delta u_{\pm}(x))^2
\geq 0$, we must have
\begin{equation}
\frac{d \theta_{\pm}}{d u_{\pm}} \geq 0,
\end{equation}
in other words, the expansion of the cross-sectional area of the
future-directed null geodesics must be locally increasing at the
throat.  This is the precise generalization of the Morris-Thorne
``flare-out'' condition to arbitrary wormhole throats. This makes
eminent good sense since the expansion is the measure of the
cross-sectional area of bundles of null geodesics, and a positive
derivative indicates that this area is locally increasing or
``flaring-out'' as one moves along the null direction. Note that
this definition is free from notions of embedding and ``shape''-functions.
So in general, we have to deal with two throats: $\Sigma_{u+}$ such
that $\theta_+ = 0$ and $d \theta_+/d u_+ \geq 0$ and $\Sigma_{u-}$
such that $\theta_{-} = 0$ and $d \theta_{-}/d u_{-} \geq 0$.  We
shall soon see that for static wormholes the two throats coalesce
and this definition automatically reduces to the static case
considered in~\cite{Hochberg-Visser,Visser-Hochberg}. The logical
development in the present paper closely parallels that of the
static case though there are many technical differences.

The conditions that a wormhole throat be both extremal and minimal
are the simplest requirements that one would want a putative throat
to satisfy and which may be summarized in the following definition
(in the following, the hypersurfaces are understood to be closed
and spatial). Since these definitions hold of course for both
throats, we momentarily drop the distinction and suppress the $\pm$
label.

\subsubsection{Definition: Simple flare-out condition}
%
{\em A two-surface satisfies the ``simple flare-out'' condition if
and only if it is extremal, $\theta=0$, and also satisfies ${d
\theta/d u} \geq 0$.}
The characterization of a generic wormhole throat in terms of the
expansion of the null geodesics shows that any two-surface satisfying
the simple flare-out condition is a {\em marginally anti-trapped
surface}, where the notion of trapped surfaces is a familiar concept
that arises primarily in the context of singularity theorems,
gravitational collapse and black hole physics~\cite{Wald,Hawking-Ellis}.
We hasten to point out however, that in the present context,
identifying a wormhole throat as a marginally anti-trapped surface
in no way, shape or form is meant to convey that we are dealing
with horizons, apparent horizons, or singularities.  Nor should
this nomenclature suggest that wormholes are somehow allied with
or are analogous to black holes or white holes.  (For some special
cases where wormholes do have applications in black hole physics
see~\cite{Visser-Hochberg}).

Generically, we would expect the inequality 
$\delta^2 A(\Sigma_{u})> 0$ to be strict, so that 
the surface is truly a minimal (not
just extremal) surface.  This will pertain provided the inequality
${d \theta/d u} > 0$ is a strict one for at least {\em some} points
on the throat.  This suggests the following definition.

\subsubsection{Definition: Strong flare-out condition}
%
{\em A two-surface satisfies the ``strong flare-out'' condition at
the point $x$ if and only if it is extremal, $\theta=0$,  satisfies
$\frac{d \theta}{d u} \geq 0$ everywhere on the surface and if at
the point $x$, the inequality is strict:}
\begin{equation}
\frac{d \theta}{d u} > 0.
\end{equation}
If the latter strict inequality holds for all $x \in \Sigma_u$ in
the surface, then the wormhole throat is seen to correspond to a
{\em strongly anti-trapped surface}. Again, this terminology
is not intended to convey any relation between wormholes and black
holes. The physical distinction between simple and strong flare-out
will become evident when we come to explore the consequences these
definitions have on the energy conditions required to maintain a
generic traversable wormhole throat.

It is sometimes sufficient and convenient to work with a weaker,
integrated form of the flare-out condition.

\subsubsection{Definition: Averaged flare-out condition}
%
{\em A two-surface satisfies the ``averaged flare-out'' condition
if and only if it is extremal, $\theta=0$, and }
\begin{equation}
\int_{\Sigma_u} \sqrt{\gamma}\, \,
{\rm sgn}
\left(\frac{d \theta}{d u}\right)  d^2x > 0,
\end{equation}
where ${\rm sgn}(x)$ is the sign of $x$. This averaged flare-out condition
places a constraint on the putative throat by asking that the
extremal surface be outward-flaring over at least half its area
before one can be justified in calling it a wormhole throat. This
definition has been carefully constructed to remain invariant under
arbitrary affine reparameterizations of the null geodesic congruence.
An apparently plausible alternative to the above, using the integral
${\cal I} \equiv \int_{\Sigma_u} \sqrt{\gamma}\; (d\theta/du)
d^2x$, is deficient in that if the integrand ${d \theta}/{d u}$
changes sign anywhere on the surface $\Sigma$ then by appropriate
affine reparameterizations of the null geodesic congruence the
integral may be made arbitrarily positive or arbitrarily
negative~\cite{Hayward}.  (Thus if one were to require the integral
${\cal I}$ to be positive for all affine parameterizations, one
would simply recover the strong flare-out condition, while if we
were to merely require that the integral ${\cal I}$ be positive
for at least one choice of affine parameterization we would have
the extremely weak constraint that ${d \theta}/{d u}$ be positive
for at least one point on the surface $\Sigma$. Either option though
mathematically consistent is physically unreasonable, and the
definition in terms of the ${\rm sgn}$ function is the best intermediate
strength definition we have found.  This comment also implies that
constraints on weighted averages of the form $\int_{\Sigma_u}
\sqrt{\gamma}\; f(x) \; (d\theta/du)  d^2x$ are too subject to
reparameterization effects to be useful.)

The conditions under which the average flare-out are appropriate
arise for example for situations with multiple throat wormholes.
Indeed, suppose we have a double throat wormhole where each of the
two throats are flared-out in the strong sense.  Then the spacetime
between the throats contains an extremal hypersurface which is not
minimal, but which can be minimal in the integrated, averaged sense.
(See, {\em e.g.}, \cite{Hochberg-Visser,Visser-Hochberg}).
Independently from this, averaged flare-out conditions of various
types crop up in energy conditions averaged over the
hypersurface~\cite{Hochberg-Visser,Visser-Hochberg}.

Finally, it is also useful to define a weighted flare-out condition.

\subsubsection{Definition: Averaged $f$-weighted flare-out condition}
%
{\em A two-surface satisfies the ``$f$-weighted flare-out'' condition
if and only if it is extremal, $\theta=0$, and }
\begin{equation}
\int_{\Sigma_u} \sqrt{\gamma}\, f(x) \,
{\rm sgn}
\left(\frac{d \theta}{d u}\right)  d^2x > 0,
\end{equation}
where $f$ is a positive definite function defined on the two-surface.

Note that the strong flare-out condition implies both the simple
flare-out condition and the averaged flare-out condition, but the
simple flare-out condition does not necessarily imply the averaged
flare-out condition (the integral might vanish). However, we see
that if the averaged $f$-weighted flare-out condition is satisfied
for all positive definite $f$, then it implies the simple flare-out
condition, which follows from identifying $f(x) = \delta u(x)^2
\geq 0$ and using the minimality constraint (\ref{E:minimality}).

\subsubsection{Technical aside: degenerate throats}

A class of wormholes for which we have to extend these definitions
arises when the wormhole throat possesses an accidental degeneracy
in the expansion of the null geodesics at the throat.  The above
discussion has been tacitly assuming that in the vicinity of the
throat we can Taylor expand the expansion
\begin{equation}
\label{E:Taylor}
\theta(x,u) = \theta(x,0) + u \left(\left.\frac{d \theta(x,u)}
{d u}\right|_{u = 0}\right) + O(u^2),
\end{equation}
with the constant term vanishing by the extremality constraint and
the first derivative term being constrained by the flare-out
conditions.

Now if the extremal two-surface has an accidental degeneracy with
the first derivative term (and possibly higher-order terms) vanishing
identically, then we would have to develop the above expansion
further out to the first non-vanishing term.  This would mean we
would have to re-phrase the flare-out in terms of these higher-order
derivatives of the null geodesic expansion.  In fact, the first
non-vanishing term would appear at odd order in $u$:
\begin{equation}
\theta(x,u) = \frac{u^{2N-1}}{(2N)!} 
\left(\left.\frac{d^{2N-1} \theta(x,u)}
{d u^{2N-1}}\right|_{u = 0}\right) + O(u^{2N}),
\end{equation}
since the surface is by definition extremal. It must be odd in $u$
otherwise the throat would be a point of inflection and not a true
minimum of the area. Simply put, even-order surface deformations
involve odd-order derivatives of the expansion. 
We can see this in another way by computing
higher-order variations in the area. The condition that it be a
minimum is
\begin{equation}
\delta^{2N}A(\Sigma_u) = \int_{\Sigma_u} \sqrt{\gamma} \,\, \,
\frac{d^{2N-1} \theta}
{d u^{2N-1}} \, (\delta u(x))^{2N}\, d^2x > 0,
\end{equation}
which leads to the flare-out condition being stated in terms of
the $({2N-1})$-th derivative of the expansion. Note: for $N=1$,
this reduces to the minimality constraint in (\ref{E:minimality}).
This motivates the following definition.

\subsubsection{Definition: $N$-fold degenerate flare-out condition}
%
{\em A two-surface satisfies the $N$-fold degenerate flare out
condition if and only if it is extremal, $\theta = 0$, the first
$(2N-2)$ $u$-derivatives of $\theta$ vanish, $(d^{2N-1} \theta(x,u)/
d u^{2N-1}) \geq 0$ everywhere on the surface and if finally, for
at least some point x on the surface, the inequality is strict:}
\begin{equation}
\frac{d^{2N-1} \theta}{d u^{2N-1}} > 0.
\end{equation}
Physically, at an $N$-fold degenerate point, the wormhole throat
is seen to be extremal up to order $2N-1$ with respect to the
derivatives of the expansion, {\em i.e.}, the flare-out condition is
delayed in the (outgoing) null direction with respect to throats
in which the flare-out occurs at $N=1$, which (by the way we have
set up the definition) corresponds to the strong flare-out condition.

These considerations bring us to the following surprising result
already alluded to above:  namely, there is no {\em a priori}
reason for the two independent null variations $\delta u_{+},\delta
u_{-}$ to single out the {\em same} minimal hypersurface.
That is, in general
\begin{equation}
\Sigma_{u+} \neq \Sigma_{u-},
\end{equation}
and we must conclude that generic time-dependent wormholes possess
two throats. If these hypersurfaces are in causal contact then it
will be possible to enter the wormhole via one throat and exit
through the other.  If the two throats are not in causal contact
then the wormhole is not two-way traversable, and you have at best
two one-way traversable wormholes with no way of getting back to
where you started from.


%
\begin{table}[htb]
\caption{Summary of the flare out conditions for wormhole
throats; all quantities are evaluated on the throat.
The flare-out conditions are understood to apply to 
both throats, and we drop the $\pm$ label.}
\vspace{0.2cm}
\begin{center}
\begin{tabular}{|l|l|l|} \hline
{\rm Flare-out condition}& 
{\rm Expansion }& {\rm Constraints on the throat} \\ \hline
   &   &   \\
{\rm simple} &$\theta = 0$& $\frac{d \theta}{d u} \geq 0$ \\ 
 &  &  \\ \hline
   &  &   \\
{\rm strong} &$\theta = 0$& $\frac{d \theta}{d u} \geq 0,$
{\rm and} $\exists x \in \Sigma_u$\,\, $\frac{d \theta}{d u} 
> 0$ 
\\ 
  &  &  \\ \hline 
  &  &   \\
{\rm strongly anti-trapped}&$\theta = 0$& $\forall x \in \Sigma_u, \,\,
\frac{d \theta}{d u} > 0$  \\
  &   &   \\ \hline 
   &     &     \\
{\rm averaged}&$\theta = 0$ & $\int_{\Sigma_u} \sqrt{\gamma}\, 
{\rm sgn} \left(\frac{d \theta}{d u}\right) d^2x > 0$ 
\\
     &    &     \\ \hline
     &    &     \\ 
{\rm $f$-averaged}&$\theta = 0$ & $\int_{\Sigma_u} \sqrt{\gamma}\, 
f(x)\, {\rm sgn}
\left(\frac{d \theta}{d u}\right) d^2x > 0$, \,\, {\rm for an}\,\,
$f(x) \geq 0$ 
\\
     &    &     \\ \hline
     &    &     \\ 
{\rm $N$-fold degenerate}&$\theta = 0$ &
$\frac{d^m \theta}{d u^m} = 0$, {\rm for}
$m = 1, \cdots , 2N-2$ \, {\rm and} 
$\frac{d^{2N-1} \theta}{d u^{2N-1}} \geq 0$  
\\
     &    &     \\
\hline
\end{tabular}
\end{center}
\end{table}
%

\subsection{Static limit}

In a static spacetime, a wormhole throat is a closed two-dimensional
spatial hypersurface of minimal area that, without loss of generality,
can be located entirely within a single constant-time spatial
slice~\cite{Hochberg-Visser,Visser-Hochberg}.  Now, for any static
spacetime, one can always decompose the spacetime metric in a
block-diagonal form as
\begin{equation}
g_{ab} = -V_a V_b + {}^{(3)}g_{ab},
\end{equation}
where $V^a = \exp[\phi](\frac{\partial}{\partial t})^a$ is a timelike
vector field orthogonal to the constant-time spatial slices and
$\phi$ is some function of the spatial coordinates only. In the
vicinity of the throat we can always set up a system of Gaussian
coordinates $n$ so that
\begin{equation}
{}^{(3)}g_{ab} = n_a n_b + \gamma_{ab},
\end{equation}
where $n^a = (\frac{\partial }{\partial n})^a$, $n^an_a = +1$, and
$\gamma_{ab}$ is the two-metric of the hypersurface.  Putting these
facts together implies that in the vicinity of any static throat
we may write the spacetime metric as
\begin{equation}
g_{ab} = -V_a V_b + n_an_b + \gamma_{ab}.
\end{equation}
But (\ref{E:decomposition}) holds in general, so comparing both metric
representations yields the identity
\begin{equation}
-l^a_{-} l^b_{+} - l^a_{+} l^b_{-} = V^a V^b + n^a n^b,
\end{equation}
and the following (linear) transformation relates the two metric
decompositions and preserves the inner-product relations in
(\ref{E:properties}):
\begin{equation}
l^a_{-} = \frac{1}{2}(V^a + n^a), \,\,\,\,
l^a_{+} = \frac{1}{2}(V^a - n^a).
\end{equation}
Since the throat is static, $\gamma_{ab}$ is time-independent,
hence when we come to vary the area (\ref{E:area-v}) with respect to
arbitrary perturbations in the two independent null directions we find that
\begin{eqnarray}
\frac{\partial \gamma_{ab}}{\partial u_{+}}\,\delta u_{+} &=& \frac{1}{2} 
\left(\exp[\phi]\frac{\partial \gamma_{ab}}{\partial t}\delta t
+ \frac{\partial \gamma_{ab}}{\partial n}\delta n\right) = \frac{1}{2} 
\frac{\partial \gamma_{ab}}{\partial n}\delta n,\nonumber \\
\frac{\partial \gamma_{ab}}{\partial u_{-}}\, \delta u_{-} &=& \frac{1}{2} 
\left(\exp[\phi]\frac{\partial \gamma_{ab}}{\partial t}\delta t
- \frac{\partial \gamma_{ab}}{\partial n}\delta n\right) = -\frac{1}{2} 
\frac{\partial \gamma_{ab}}{\partial n}\delta n.
\end{eqnarray}
Thus the variations are no longer independent, and reduce to
taking a single surface variation in the spatial Gaussian direction.
So, $\theta_+ = 0 \Longleftrightarrow \theta_{-} = 0$ at the same
hypersurface, proving that $\Sigma_{u_+} = \Sigma_{u_-}$ in the
static limit, and so static wormholes have only one throat.
An exhaustive analysis of the geometric structure of the generic
static traversable wormhole may be found
in~\cite{Hochberg-Visser,Visser-Hochberg}.

\section{Constraints on the stress-energy}

With the definition of wormhole throat made precise we now turn to
derive constraints that the stress energy tensor must obey on (or
near) any wormhole throat. The constraints follow from combining
the Raychaudhuri equation (\ref{E:Raychaudhuri}) with the flare-out
conditions, and using the Einstein equation (\ref{E:Einstein}).
It is clear that these constraints apply with equal validity at
both the $+$ and $-$ throats, and in the following we cover both
classes simultaneously and without risk of confusion by dropping
the $\pm$-labels. We first treat the zero-torsion case.

\subsection{Zero torsion}

Since all throats are extremal hypersurfaces $(\theta = 0)$ 
the Raychaudhuri equation at the throat
(\ref{E:Raychaudhuri}) reduces to
\begin{equation}
\label{E:vthroat}
\frac{d\theta}{d u} + \sigma^{ab}\sigma_{ab} = -8\pi T_{ab} \; l^al^b,
\end{equation}
where we have used the Einstein equation (\ref{E:Einstein}) after
setting the torsion terms to zero and the fact that the 
null geodesic congruences are hypersurface orthogonal, so that the
twist $\omega_{ab} = 0$ vanishes identically on the throat.  We
make no claim regarding the shear, except to point out that since
$\sigma_{ab}$ is purely spatial, its square $\sigma^{ab}\sigma_{ab}
\geq 0$ is positive semi-definite everywhere (not just on the throat).
Consider a marginally anti-trapped surface, {\em i.e.}, a throat
satisfying the simple flare-out condition. Then the stress energy
tensor on the throat must satisfy
\begin{equation}
T_{ab} \; l^al^b \leq 0.
\end{equation}
The NEC is therefore either violated, or on the verge of being
violated ($T_{ab} \; l^al^b \equiv 0$), on the throat.  Of course,
whichever one of the two null geodesic congruences ($l_{+}$ or
$l_{-}$) you are using to define the wormhole throat (anti-trapped
surface), you must use the {\em same} null geodesic congruence for
deducing null energy condition violations.

For throats satisfying the strong flare-out condition, we have
instead the stronger statement that for all points on the
throat,
\begin{equation}
T_{ab}l^al^b \leq 0, \,\,{\rm and}\,\,
\exists x \in \Sigma_u \,\,\,{\rm such}\,{\rm that}\,\,\, 
T_{ab} \; l^a l^b < 0,
\end{equation}
so that the NEC is indeed violated for at least {\em some} points
lying on the throat.  By continuity, if $T_{ab} \; l^al^b < 0$ at $x$,
then it is strictly negative within a finite open neighborhood of
$x$: $B_{\epsilon}(x)$.  For throats that are strongly anti-trapped
surfaces, we derive the most stringent constraint stating that
\begin{equation}
T_{ab}\; l^a l^b < 0 \,\,\, \forall x \in \Sigma_u,
\end{equation}
so that the NEC is violated {\em everywhere} on the throat.

Weaker, integrated energy conditions are obtained for throats
satisfying the averaged flare-out conditions. For a throat that is
flared-out on the average, integrating the Raychaudhuri equation
(\ref{E:vthroat}) over the throat implies
\begin{equation}
\int_{\Sigma_u} \sqrt{\gamma} \, {\rm sgn}( T_{ab}\; l^a l^b ) \, d^2 x < 0,
\end{equation}
indicating that the NEC, when averaged over the throat, is strictly
violated (Warning: this has nothing to do with the violation of
the averaged null energy condition, or ANEC. In the ANEC, the
averaging is defined to take place along inextendible null geodesics.
See in particular~\cite{FSW}.)  By the same token, throats satisfying
the $f$-weighted averaged flare-out condition imply that
\begin{equation}
\int_{\Sigma_u} \sqrt{\gamma} \,f(x)\, {\rm sgn}(
T_{ab} \; l^a l^b) \, d^2 x < 0,
\end{equation}
indicating that the sign of the NEC, when weighted with the positive
definite function $f(x)$ is strictly violated on the average over
the throat.

What can we say about the energy conditions in the region surrounding
the throat? This requires knowledge of the expansion, shear and
twist in the neighborhood of the throat. Luckily, we can dispense
with the twist immediately. Indeed, the (torsion-free) twist equation
(\ref{E:vtwist}) is a simple, first-order linear differential equation:
\begin{equation}
\frac{d \omega_{ba}}{d u} = -\theta \omega_{ba}
-2\sigma^c_{[a}\omega_{b]c},
\end{equation} 
whose exact solution (if somewhat formal in appearance) is
\begin{equation}
\omega_{ab}(u) = \exp\left(-\int_0^u \theta(s) ds \right) \;
{\cal U}_a{}^c(u) \; {\cal U}_b{}^d(u) \omega_{cd}(0),
\end{equation}
where the quantity ${\cal U}(u)$ denotes the path-ordered exponential
\begin{equation}
{\cal U}_a{}^c(u) = {\cal P}\exp\left(-\int_0^u \sigma \; ds\right){}_a{}^c.
\end{equation}
So, an initially hypersurface orthogonal congruence remains twist-free
everywhere, both on and off the throat:  $\omega_{ba}(0) = 0
\Rightarrow \omega_{ba}(u) = 0$.  Then the equation
\begin{equation}
\frac{d\theta}{d u} + \frac{1}{2}\theta^2 +
\sigma^{ab}\sigma_{ab} = -8\pi T_{ab}\; l^a l^b,
\end{equation}
is seen to be valid for all $u$. Coming back to simply-flared
throats, we have two pieces of information regarding the expansion:
namely that $\theta(0) = 0$ and $(d \theta(u)/d u)_{u=0} \geq 0$,
so that if we expand $\theta$ in a neighborhood of the throat as
in (\ref{E:Taylor}), then we have that
\begin{equation}
\frac{d \theta(u)}{d u} = 
\left. \frac{d \theta(u)}{d u}\right|_{u=0} + O(u),
\end{equation}
so over each point $x$ on the throat, there exists a finite range
in affine parameter $u \in (0,u^*_x)$ for which $\frac{d \theta(u)}{d
u} \geq 0$. Since both $\theta^2$ and $\sigma^{ab}\sigma_{ab}$ are
positive semi-definite, we conclude that the stress-energy is either
violating, or on the verge of violating, the NEC along the partial
null curve $\{x\}\times(0,u^*_x)$ based at $x$.  If the throat is
of the strongly-flared variety, then we see that the NEC is definitely
violated at least over some finite regions surrounding the throat:
$\bigcup_x \{x\}\times(0,u^*_x)$, and including the base points
$x$.  For strongly anti-trapped surfaces, the NEC is violated everywhere
in a finite region surrounding the entire throat, and including
the throat itself.

Finally, if the throat is $N$-fold degenerate (and $N>1$), then
there exist points $x$ on the throat for which $(d^{2N-1}\theta(x,u)/
d u^{2N-1})|_{u=0} > 0$. This implies that the first derivative
\begin{equation}
\frac{d \theta(x,u)}{d u} =
\left. \frac{(2N-1)u^{2N-2}}{(2N)!}\frac{d^{2N-1} \theta(x,u)}
{d u^{2N-1}}\right|_{u=0} + O(u^{2N-1}),
\end{equation}
is positive along a partial null curve $\{x\}\times(0,u^*_x)$ based
at $x$ and it follows by (\ref{E:vthroat}) that the NEC is violated
along the finite ``bristles'' $\bigcup_x \{x\}\times(0,u^*_x)$.

\subsection{Non-zero H-torsion }

Torsion, although contributing additional terms to the Einstein
(\ref{E:Einstein}) and Raychaudhuri equations (\ref{E:Raychaudhuri})
does not necessarily alleviate the problem of the violation of the
NEC on or near wormhole throats. This state-of-affairs holds at
both throats so without loss of generality, take $(+)$-throats and
consider the term linear in $H$ that appears in (\ref{E:Raychaudhuri}).
This can be simplified as follows:
\begin{eqnarray} 
l^c_{+} \; H^d_{cb} \; {B^{+}}^b_d &=& 
l^c_{+} \; H^d_{cb}\left(v_d^{+b} - l^b_{+} \gamma_d^e l^c_{-}
\nabla_el_{+ c}\right),\nonumber \\
 &=& 
l^c_{+} \; H^d_{cb} \; \omega_d^{+ b},
\end{eqnarray}
since the mixed spatial-null components of 
${B^{+}}^b_d$ are orthogonal
to $H^d_{cb}$, and by virtue of the latter's antisymmetry, projects
out the twist from the purely spatial tensor $v^{+b}_d$.  Now consider
an initially hypersurface orthogonal null congruence, then at the
throat of the wormhole we have
\begin{equation}
\label{E:torsion}
\frac{d\theta_{+}}{d u_{+}} + {\sigma^{+}}^{ab}{\sigma^{+}}_{ab} = 
-8\pi T_{ab} \; l^a_{+}l^b_{+}
-2H_{ade}H_b^{de} \; l^a_{+} l^b_{+},
\end{equation}
after using the expression for the Ricci tensor in (\ref{E:Ricci}).  

We could now run through the list of flare-out conditions (see
Table I) as before and we would obtain, as expected, constraints
on the combination of stress-energy and torsion appearing on the
right hand side of (\ref{E:torsion}).  Thus, for a simply-flared
throat, or marginally anti-trapped surface, we must have
\begin{equation}
\label{E:ineq}
4\pi T_{ab} \; l^a_{+} l^b_{+} +H_{ade}H_b^{de} \; l^a_{+} l^b_{+} \leq 0,
\end{equation}
at the throat and one might propose sweeping the violations into
the torsion sector.  We will find that this is not possible.  For
illustrative purposes, suppose we consider the ansatz
\begin{equation}
H_{abc} = \frac{1}{\sqrt{-g}}\epsilon_{abce}w^e(x),
\end{equation}
for any vector field $w^e$.  Then the combination
\begin{equation}
H_{ade}H_b^{de} \; l^a_{+} l^b_{+} = +2(w^al_{+ a})^2 \geq 0,
\end{equation}
is positive definite for all $w$. Such a torsion-field aggravates
the violation of the NEC and all of the above constraints on the
stress-tensor derived at and near the throat in the zero-torsion
case apply as well to throats in the presence of this class of
non-zero torsion.  Actually, with a little more work, it is 
possible to relax the assumption of total antisymmetry 
and demonstrate 
that {\em all} torsion leads to enhanced
violation of the NEC! To see why this comes about first consider
the general decomposition of an arbitrary antisymmetric rank-two
tensor $A_{ab} = - A_{ba}$ in terms of null vectors and spatial
projector. We find that we can write
\begin{eqnarray}
A_{ab} &=& a l_{-[a}l_{+b]} + \gamma_{[a}^c \gamma_{b]}^d A_{cd}
\nonumber
\\
&+& 2l_{-[a}l^c_{+}\gamma_{b]}^d A_{dc}
 + 2 l_{+[a} l^c_{-}\gamma_{b]}^d A_{dc},
\end{eqnarray}
where the coefficient $a = -2l^c_{-} l^d_{+} A_{cd}$.  Now evaluate
this for $A_{de} = l^a_{+} H_{ade}$. One finds that $a =
-2l^a_{-}l^b_{+}l^c_{+} H_{cab} = 0$. The third term above also
vanishes since $l_{-[a}\gamma_{b]}^d l^c_{+} l^e_{+} H_{edc} = 0$,
which leaves us with
\begin{equation}
\label{E:A}
A_{ab} = \tilde A_{ab} + 2l_{+[a}l^c_{-}\gamma_{b]}^d A_{cd},
\end{equation}
where $\tilde A_{ab} = \gamma_{[a}^c \gamma_{b]}^d A_{dc}$ is a
purely spatial tensor.  Now, the square of (\ref{E:A}) involves only
the purely spatial components:
\begin{equation}
A_{de}A^{de} \equiv l^a_{+}H_{ade} l^b_{+}H_{b}^{de} = 
\tilde A_{de}\tilde A^{de} \geq 0,
\end{equation}
and this is precisely the combination appearing in  (\ref{E:ineq}).
Thus, the torsion terms cannot be made to absorb any energy
violations.  On the contrary, torsion tends to focus null geodesics.
While the ``normal'' stress-energy must continue to violate (or be
on the verge of violating) the NEC on the throat, the presence of
any non-zero torsion  does act to lessen the violation off the
throat. This is simply because torsion acts as a source of twist,
and even if the twist vanishes on the throat, nonvanishing twist
is eventually generated in the neighborhood surrounding the throat,
as can be appreciated by examining Eq. (\ref{E:vtwist}), and twist
comes in with the just the right sign in the Raychaudhuri equation.
Of course, without further input, we have no way of knowing if this
happens in the region near the throat or far away from the throat.
If it occurs near the throat, then the energy violations in that
region might be (partially) absorbed into the twist, but
the violation persists nonetheless.

\section{Worked examples}
\subsection{Conformally expanding Morris-Thorne wormhole}

We shall illustrate these basic concepts and constructs with the
following explicit example.  Consider the time-dependent spherically
symmetric inter-Universe wormhole described by a pair of coordinate
patches in which the metric takes the form
\begin{equation}
\label{E:kar}
ds^2 = 
\Omega^2(t)\left( 
-dt^2 + \frac{dr_{1,2}^2}{1-\frac{b(r_{1,2})}{r_{1,2}}} +
r_{1,2}^2 [d\theta^2 + \sin^2 \theta \, d\phi^2] \right).
\end{equation}
This metric is conformally related to a zero-tidal force inter-Universe
Morris-Thorne wormhole by a simple time-dependent but space-independent
conformal factor~\cite{Roman,Kar1,Kar2}.  (Other versions of
time-dependent wormholes are discussed
in~\cite{Wang-Letelier,Kim,Time-dependent}.) Each coordinate system
used to exhibit the metric given above covers only half the wormhole
spacetime, and there are two radial coordinates, $r_1$ and $r_2$,
each of which runs only from $r_0$ to infinity, where $r_0$ is
obtained by solving the implicit equation $b(r_0) = r_0$.
See~\cite{Morris-Thorne,Visser-Book}. The two radial coordinates
cover two distinct universes and overlap only at $r_1 = r_0 = r_2$
which defines the {\it center} of the wormhole (we will find that
the center coincides with the throat only in the static limit).
For simplicity this wormhole is taken to be symmetric under
interchange of the two asymptotically flat regions but this is not
essential to the analysis.

It should be clear that we look for throats within {\it each}
coordinate patch separately.  We will see below that for suitable
energy conditions, the above metric corresponds to a wormhole with
two time-dependent throats, each throat residing in one of the two
Universes joined by the wormhole.

\subsubsection{First coordinate patch}

The throats, when and if they exist, will be located on spheres of
(instantaneous) radii $\Omega(t) r_1$ (where $r_1 \geq r_0$) possessing
the spatial metric (written in block-diagonal form)
\begin{equation}
\label{E:sphere}
\gamma_{1{}ab} = \Omega^2 \; r_1^2 \; 
\left( \matrix{ {\bf 0} & {\bf 0} \cr
                {\bf 0} & \left[ \matrix{ 1 & 0 \cr
                                            0 & \sin^2 \theta} \right] }
\right).
\end{equation}
We can easily find a set of two independent null vectors orthogonal to the
spheres in this patch; they are given by
\begin{equation}
l^a_{\pm} = \frac{1}{\sqrt{2}\Omega}
\left(1, \pm \left(1-\frac{b(r_1)}{r_1}\right)^{\frac{1}{2}},0,0\right),
\end{equation}
and it is easy to verify that all the inner-product relations
(\ref{E:properties}) are satisfied and that the metric (\ref{E:kar}) in this
patch
can be decomposed in terms of $l_{+a},l_{-a},\gamma_{1{}ab}$ just as in
(\ref{E:decomposition}). 
The expansions of these null geodesics are calculated 
in a straightforward manner:
\begin{equation}
\theta_\pm = \gamma_1^{ab}\nabla_al_{\pm b}=
\sqrt{2}\frac{\dot \Omega}{\Omega^2} \pm 
\frac{\sqrt{2}}{r_1\Omega}\left(1 - \frac{b(r_1)}{r_1}\right)^{\frac{1}{2}},
\end{equation}
where the overdot stands for the derivative with respect to
(conformal) time $t$.
The derivatives, taken with respect to the affine parameter, used
for testing for flare-out are $(d\theta_{+}/d u_{+} = l^t_+ \partial
\theta_{+}/\partial t + l^r_+ \partial \theta_{+}/\partial r)$,
{\em etc.},
\begin{eqnarray}
\frac{d\theta_\pm}{d u_\pm} &=& \frac{1}{\Omega^2}
\left( 
\left(\frac{\ddot \Omega}{\Omega}- 2\frac{{\dot \Omega}^2}{\Omega^2}\right) 
\mp
\frac{\dot \Omega}{r_1\Omega}
\left(1 - \frac{b(r_1)}{r_1}\right)^{\frac{1}{2}}
 - \frac{1}{r_1^2}\left(1 - \frac{b(r_1)}{r_1}\right) \right. \nonumber \\
&+&  \left. \frac{1}{2r_1^2} \left(-b'(r_1) + \frac{b(r_1)}{r_1}\right) 
\right).
\end{eqnarray}

Now we can search for throats in this patch. First we locate the
extremal hypersurfaces; these coincide with the zeroes of the
expansions:
\begin{equation}
\theta_{\pm} = 0 \Longleftrightarrow 
\frac{1}{r_1}\left(1 - \frac{b(r_1)}{r_1} \right)^{\frac{1}{2}} 
= \mp \frac{\dot \Omega}{\Omega},
\end{equation}
which defines the time-dependent throat radius $r_1^*(t)$ implicitly.
We note that the factor involving the square-root is always positive
semi-definite, hence we find that (in the $r_1$  coordinate patch)
it is only $\theta_{-}$ that can vanish for an expanding ($\dot
\Omega > 0$) background while it is $\theta_{+}$ that can vanish
for a collapsing ($\dot \Omega < 0$) background. There is therefore,
always only one extremal hypersurface in the first patch.

Irrespective of expansion or collapse, the flare-out evaluated
on that extremal hypersurface works out to be  
\begin{equation}
\label{E:flare}
\left.\frac{d \theta_{\pm}}{d u_{\pm}}\right|_{\theta_{\pm}= 0} =
\frac{1}{\Omega^2}
\left( 
\left[
\frac{\ddot \Omega}{\Omega}-2\frac{{\dot \Omega}^2}{\Omega^2}
\right]
+ \frac{1}{2r_1^*(t)^2}\left[
-b'(r_1^*(t)) + \frac{b(r_1^*(t))}{r_1^*(t)}
\right]
\right).
\end{equation}
The flare-out of the hypersurface is a function of time.  Note that
the second grouped term on the right hand side is always greater
than or equal to zero while the first grouped term can in principle,
have any sign, depending on the nature of the background expansion
(or contraction). This observation was first proposed in~\cite{Kar1,Kar2}
as a means of temporarily suspending the energy condition violations
for dynamic wormholes.  However, the Einstein tensor associated
with the above metric (\ref{E:kar}) can be easily worked
out~\cite{Kar1,Kar2} and taking its projection along the radial
null direction yields the combination
\begin{equation}
G_{\hat t\hat t} + G_{\hat r \hat r} 
= 8\pi(\rho_1 - \tau_1)=
\Omega^{-2} 
\left( 
-{b(r_1)\over r_1^3} + {b'(r_1)\over r_1^2}   
- 2 {\ddot\Omega\over\Omega} + 4 {\dot\Omega^2\over\Omega^2} 
\right),
\end{equation}
where $\rho_1$ and $\tau_1$ denote the energy density and radial
tension as seen by an observer in the proper reference frame.
Evaluate this at $r_1=r_1^*(t)$ and compare it to (\ref{E:flare}) to
conclude that any conformal factor that is chosen so as to suspend
the violation of the NEC, will at the same time eradicate
the flare-out condition:
\begin{equation}
(\rho_1 - \tau_1) \geq 0  \Longleftrightarrow 
\frac{d \theta_{\pm}}{d u_{\pm}}|_{r_1^*(t)}  \leq 0,
\end{equation}
and the hypersurface at $r_1^*(t)$ will {\em not be flared-out!}
In other words, the extremal hypersurface will be a throat of
the simply flared-out variety if and only if the NEC is violated
or on the verge of being violated there.

This is completely compatible with the topological censorship
theorem~\cite{FSW}. If one picks an ingoing radial null geodesic
along which the NEC is always satisfied, then by the above argument
the expansion can never flare out, one is forced to continue moving
inward,  and so one cannot pass through a wormhole throat.

\subsubsection{Second coordinate patch}

Many of the results from the first coordinate patch can be carried
over to the second coordinate patch with a few key flips in signs.

The throats in this second patch, when and if they exist, will be
located on spheres of (instantaneous) radii $\Omega(t) r_2$ (with
$r_2 \geq r_0$) possessing the spatial metric (written in block-diagonal
form)
\begin{equation}
\label{E:sphere2}
\gamma_{2{}ab} = \Omega^2 \; r_2^2 \; 
       \left( \matrix{ {\bf 0} & {\bf 0} \cr
                       {\bf 0} & \left[ \matrix{ 1 & 0 \cr
                                            0 & \sin^2 \theta} \right] }
\right).
\end{equation}
We can easily find a set of two independent null vectors orthogonal to the
spheres in this patch; they are given by
\begin{equation}
l^a_{\pm} = \frac{1}{\sqrt{2}\Omega}
\left(1, \mp \left(1-\frac{b(r_2)}{r_2}\right)^{\frac{1}{2}},0,0\right).
\end{equation}
It is easy to verify that the key sign flip above guarantees that
the vector fields $l_\pm$ defined patch one connect smoothly with
their definitions on patch two. Furthermore all the inner-product
relations (\ref{E:properties}) are satisfied and that the metric
(\ref{E:kar}) in this patch can still be decomposed in terms of
$l_{+a}$, $l_{-a}$, and $\gamma_{2{}ab}$ just as in
(\ref{E:decomposition}).
Their respective expansions are calculated in a straightforward manner:
\begin{equation}
\theta_\pm = \gamma_2^{ab}\nabla_a l_{\pm b}=
\sqrt{2}\frac{\dot \Omega}{\Omega^2} \mp 
\frac{\sqrt{2}}{r_2\Omega}\left(1 - \frac{b(r_2)}{r_2}\right)^{\frac{1}{2}}.
\end{equation}
The search for throats in this second patch proceeds just as above.
For the location of the extremal hypersurfaces we now have
\begin{equation}
\theta_{\pm} = 0 \Longleftrightarrow 
\frac{1}{r_2}\left(1 - \frac{b(r_2)}{r_2} \right)^{\frac{1}{2}} 
= \pm \frac{\dot \Omega}{
\Omega},
\end{equation}
which now defines the throat radius $r_2^*(t)$ implicitly.  We
again note that the left hand side is always positive semi-definite,
hence we find that it is now $\theta_{+}$ that vanishes for an
expanding background while it is $\theta_{-}$ that vanishes for a
collapsing background (in this patch!). Therefore, there is again
exactly one extremal hypersurface in this patch.  Note that because
of the crucial sign flip, whichever of the two expansions it is
that vanishes in coordinate patch one, it is the {\em other}
expansion that will now vanish in patch two.

Because of the assumed symmetry between the two patches the rest
of the analysis follows through without difficulty and we can again
see that any conformal factor $\Omega$ that is chosen so as to
suspend the violation of the NEC, will at the same {\em time}
eradicate the flare-out condition at this second throat:
\begin{equation}
(\rho_2 - \tau_2) \geq 0  \Longleftrightarrow 
\frac{d \theta_{\pm}}{d v_{\pm}}|_{r_2^*(t)}  \leq 0.
\end{equation}
Once again, this extremal hypersurface will be a throat of
the simply flared-out variety if and only if the NEC is violated
or on the verge of being violated there.

(As indicated previously, the assumption that the wormhole is
symmetric under interchange of the two asymptotically flat regions
is not essential to the analysis. To generalise this point one just
needs to choose two un-equal shape functions $b_1(r_1)$ and $b_2(r_2)$
that need be linked only by the fact that they simultaneously
satisfy $b_1(r_0)=r_0=b_2(r_0)$. It is now a simple exercise to go
through the preceding formulae making minor changes as appropriate.)

\subsubsection{Static limit}

In the static limit, we have $\dot \Omega = 0$ and the simultaneous
vanishing of the expansions now occurs at the unique point where
the two coordinate patches overlap:  $b(r_0) = r_0$, this value
being none other than the center of the wormhole: therefore, the
static wormhole has only one throat, and the throat coincides with
the center of the wormhole.  We thus recover the zero-tidal force
Morris-Thorne wormhole.  Reality of the expansions further restrains
the $b$-function to satisfy $b(r) \leq r$ so that $b'(r_0) \leq
1$.  The flare-outs of this unique throat with respect to either
coordinate patch are
\begin{equation}
\left.\frac{d \theta_{\pm}}{d u_{\pm}}\right|_{r_0} =
\frac{1}{r_0^2}\left(
-b'(r_0) + 1\right) \geq 0,
\end{equation}
so that the sphere of constant radius $r_0$ is a throat satisfying
the simple flare-out condition and is therefore a marginally
anti-trapped surface.  It follows immediately from the above
theorems, and in complete agreement with the standard analyses,
that the NEC is either violated, or on the verge of being violated,
at the throat.  Note of course, that if these inequalities are
strictly positive at any point on the throat, then these derivatives
are strictly positive everywhere on the the throat (by spherical
symmetry) and the throat satisfies the strong flare-out condition
everywhere and is therefore a strongly anti-trapped surface. The
NEC is strictly violated in this case.

\subsubsection{Summary}

This worked example shows how important it is to distinguish the
``center'' of the wormhole, defined by looking at the spatial
behaviour of a fixed time-slice, from the throat of the wormhole,
defined by the flare-out condition applied to null geodesics that
are actually trying to traverse the wormhole.

If the null geodesics ever succeed in getting through the traversable
wormhole, into the ``other universe'', then they must at some stage
have passed a region where their expansion satisfied the flare-out
condition,  and this region is what we define to be the throat of
the wormhole. By the analysis of this paper, the NEC must be violated
at or near this throat. The ``center'' of the wormhole is the wrong
place to look for NEC violations, except in the static limit where
the two throats coalesce trapping the center between them.

\subsection{\protect{General time-dependent spherically symmetric\hfil\break
traversable wormhole}}

The most general metric describing a time-dependent spherically
symmetric spacetime can (with appropriate choice of an atlas of
coordinate patches) be written as
\begin{equation}
ds^2 =  -e^{2\psi}\left(1 - \frac{2m}{r}\right)dv^2
+ 2 e^{\psi} dv\,dr + r^2(d\theta^2 + \sin^2 \theta d\phi^2).
\end{equation}
Here $\psi(v,r)$ and $m(v,r)$ are two independent functions of the
radial coordinate $r$ and an advanced time-parameter $v$ ($ v
\approx t + r $ at large $r$)~\cite{Bardeen}.  This metric can also
be adapted to describe an inter-Universe wormhole. As in the previous
example, the coordinate system employed covers only half the wormhole
spacetime and so two patches will be required and the radial
coordinate $r \in [r_0,\infty)$, where $r_0(v)$ is again the center
of the wormhole. We should then introduce four independent functions:
$\psi_{1,2}$ and $m_{1,2}$ where the labels refer to the two
coordinate patches. These functions must satisfy a smoothness
condition at $r=r_0(v)$ if there is to be no delta function material
concentrated on the throat (the extrinsic curvatures should match
across the  center of the wormhole, see~\cite{Visser-Book}).

In the interest of brevity and notational economy, we will focus
on one of the two coordinate patches only. So consider one of the
Universes joined by the wormhole.  A throat, when it exists, will
be a sphere of radius $r \geq r_0$ with spatial metric given by
(\ref{E:sphere}) with $\Omega = 1$. The two independent sets of null
vectors orthogonal to the sphere are found to be given by
\begin{equation}
l^a_{+} = 
\left(1,\frac{1}{2}e^{\psi}\left(1-\frac{2m}{r}\right),0,0\right), 
\,\,\,
l^a_{-} = (0,-e^{-\psi},0,0).
\end{equation}
The expansions of the associated two sets of null rays are 
\begin{equation}
\theta_{+} = 2\gamma^{\theta \theta}\nabla_{\theta}l_{+\theta} =
\frac{1}{2} e^{\psi}\left(1-\frac{2m}{r}\right),
\end{equation}
and
\begin{equation}
\theta_{-} = 2\gamma^{\theta \theta}\nabla_{\theta}l_{-\theta}=
-\frac{2}{r} e^{-\psi},
\end{equation}
respectively.
Provided $\psi(r,v)$ is non-singular (a good idea if there are to
be no horizons!), the only expansion which can have zeros is
$\theta_{+}$ and $\theta_{+} = 0 \Longleftrightarrow 2m(r,v) = r$,
so that $r= r(v)$ gives the time-dependent radius of the extremal
sphere.

The flare-out evaluated at this hypersurface is readily calculated
to be $(\frac{d}{du_+} = l^a_{+}\nabla_a = l^v_{+}\frac{\partial
}{\partial v} +l^r_{+} \frac{\partial }{\partial r})$
\begin{equation}
\left.\frac{d\theta_{+}}{du_{+}}\right|_{\theta_{+} = 0} = 
-\frac{2}{r^2(v)}e^{\psi}
\left. \left(\frac{\partial m(r,v)}{\partial v} \right)\right|_{r(v)}.
\end{equation}
The Einstein equations are easy to work out in this metric. At this
throat of the wormhole, the null-null component yields
\begin{equation}
\left. \left(\frac{\partial m(r,v)}{\partial v} \right)
\right|_{\theta_{+} = 0} =
4\pi r^2 T_{vv},
\end{equation}
so that 
\begin{equation}
\left.\frac{d\theta_{+}}{du_{+}}\right|_{\theta_{+} = 0} \geq 0 
\Longleftrightarrow \left. T_{vv}\right|_{\theta_{+} = 0} \leq 0.
\end{equation}
Once again, this throat will be simply-flared if and only if the
null energy condition is violated, or on the verge of being violated,
at the throat. If the violations are suspended at the throat, the
hypersurface will not satisfy any flare-out condition, and so ceases
to be a throat. (For instance, this is what occurs in
Refs.~\cite{Kar1,Kar2,Wang-Letelier,Kim,Time-dependent}.) An entirely
similar analysis can be carried out for the other coordinate patch.
Again, there are are total of two time dependent throats and again,
they coalesce into a single throat located at $r_0$ in the static
limit.

\section{Discussion}

We have presented a local geometric definition of a wormhole throat
that generalizes the notion of ``flare-out'' to an arbitrary
time-dependent wormhole and is free from technical assumptions about
global properties.  Flare-out is manifested in the properties of
light rays (null geodesics) that traverse a wormhole: bundles of
light rays that enter the wormhole at one mouth and exit from the
other must have cross-sectional area that first decreases, reaching
a true minimum at the throat, and then increases. These properties
can be quantified precisely in terms of the expansion $\theta_\pm$
of the (future-directed) null geodesics together with its derivative
$d \theta_{\pm}/d u_{\pm}$, where all quantities are evaluated at
the two-dimensional spatial hypersurface comprising the throat.
Strictly speaking, this flaring-out behavior of the outgoing null
geodesics ($l_{+}$) defines one throat: the ``outgoing'' throat.
But one can also ask for the flaring-out property to be manifested
in the propagation of the set of ingoing null geodesics ($l_{-}$)
as they traverse the wormhole, and this leads one to define a
second, or ``ingoing'' throat. In general, these two throats need
not be identical, but for the static limit they do coalesce and
are indistinguishable.

The flaring-out property implies that all wormhole throats are in
fact {\em anti-trapped} surfaces, an identification that was
anticipated some time ago by Page~\cite{Page}. With this definition
and using the Raychaudhuri equation, we are able to place rigorous
constraints on the Ricci tensor and the stress-energy tensor at
the throat(s) of the wormhole as well as in the regions near the
throat(s). We find, as expected, that wormhole throats generically
violate the null energy condition and we have provided several
theorems regarding this matter.

The nature of the energy-condition violations associated with
wormhole throats has led numerous authors to try to find ways of
evading or minimizing the violations.  Most attempts to do so focus
on alternative gravity theories in which one may be able to force
the extra degrees of freedom to absorb the energy-condition violations
(some of these scenarios are discussed in~\cite{Visser-Hochberg},
see also \cite{Brans-Dicke-Papers,Kar3}). But the energy condition
violations are still always present, as sweeping the energy condition
violations into a particular sector surely does not make the problem
go away.  As a striking case in point, we have treated in detail
the case of gravity plus torsion. If we identify the torsion with
that appearing naturally in the spectrum of closed strings, then
we find it actually worsens the violations of the NEC at the throats.
More recently it has been realized that time-dependence lets one
move the energy condition violating regions around in $\em
time$~\cite{Kar1,Kar2,Wang-Letelier,Kim,Time-dependent}.  Temporary
suspension of the violation of the NEC at a time-dependent throat
also leads to a simultaneous obliteration of the flare-out property
of the throat itself, so this strategy ends up destroying the throat
and nothing is to be gained. (See also~\cite{Visser-Hochberg}.) In
arriving at this conclusion it is crucial to note that we have
defined flare-out in terms of the expansion properties of light
rays at the throat and {\em not} in terms of ``shape'' functions
or embedding diagrams.  While the latter can certainly be used
without risk for detecting flare-out in static wormholes, they are
at best misleading if applied to dynamic wormholes. This is simply
because the embedding of a wormhole spacetime requires selecting
and lifting out a particular time-slice and embedding this
instantaneous spatial three-geometry in a flat Euclidean ${\bf
R}^3$. For a static wormhole, any constant time-slice will suffice,
and if the embedded surface is flared-out in the spatial direction
orthogonal to the throat, then it is flared-out in spacetime as
well. But if the wormhole is dynamic, flare-out in the spatial
direction does not imply flare-out in the {\em null} directions
orthogonal to the throat.

\section*{Acknowledgements}

The work of M.V. was supported by the US Department of Energy.
Additionally, M.V. wishes to thank the members of LAEFF (Spain)
for their hospitality during early phases of this work and acknowledges
interesting and constructive comments made by Sean Hayward~\cite{Hayward}.


\end{document}